\documentclass[sigchi]{acmart}
\setcopyright{none} % Remove copyright block
\acmSubmissionID{} % Remove submission ID
\acmDOI{} % Remove DOI
\acmISBN{}
\acmYear{} % Remove the year
\copyrightyear{} 
\acmSubmissionID{}
\acmConference[Arxiv]{Online}{2025} % Clear conference name, location, and date

\usepackage{subfig}
\AtBeginDocument{%
  \providecommand\BibTeX{{%
    \normalfont B\kern-0.5em{\scshape i\kern-0.25em b}\kern-0.8em\TeX}}}

\begin{document}

\title{Pneutouch: Exploring the affordances and interactions of haptic inflatables through a wrist-worn interface}

\author{Frank Wencheng Liu*, Mason Manetta}
\authornote{Both authors contributed equally to this work.}
\author{Prasad Borkar, Byron Lahey, Assegid Kidane, Robert LiKamWa}
\affiliation{
  \institution{Arizona State University}
  \city{Tempe}
  \country{United States}
}

\renewcommand{\shortauthors}{Liu et al.}

\begin{abstract}
Haptic sensations that align with virtual reality (VR) experiences have a profound impact on presence and enjoyment. There is potential to explore the dynamic capabilities of pneumatic inflatables to offer immersive sensations in virtual environments, including variations in shape, size, and stiffness. We introduce Pneutouch, an ungrounded and untethered wrist-worn device designed as a pneumatic haptic interface for VR interactions. Pneutouch's dynamic inflation ability enables programmable stiffness and shape change of haptic proxies. Additionally, multiple haptic proxies can be delivered into and out of the user's hand grasp. 
We describe the implementation of the Pneutouch device. We conducted user studies to demonstrate the affordances of pneumatic inflatables and assessed the device's efficacy in providing haptic feedback.
With Pneutouch, our goal is to expand what can be touched in the virtual space and bring more immersion into virtual reality.
\end{abstract}

\begin{CCSXML}
<ccs2012>
   <concept>
       <concept_id>10010583.10010588.10010598.10011752</concept_id>
       <concept_desc>Hardware~Haptic devices</concept_desc>
       <concept_significance>300</concept_significance>
       </concept>
   <concept>
       <concept_id>10003120.10003121.10003124.10010866</concept_id>
       <concept_desc>Human-centered computing~Virtual reality</concept_desc>
       <concept_significance>300</concept_significance>
       </concept>
 </ccs2012>
\end{CCSXML}

\ccsdesc[300]{Hardware~Haptic devices}
\ccsdesc[300]{Human-centered computing~Virtual reality}

\keywords{HCI, haptics, pneumatics, virtual reality}

\maketitle

\section{Introduction}

Haptic technologies, designed to provide tangible sensations to users \cite{canitouchthis}, have become integral in augmenting virtual reality (VR) experiences. The value of haptics in VR lies in their ability to bridge the gap between the virtual and physical worlds, providing users with sensory feedback that enhances immersion, realism, and interaction fidelity. Without haptic feedback, VR experiences can feel incomplete, as the lack of tactile information makes virtual objects and environments less convincing and engaging. Haptic feedback is not simply a binary condition that indicates physical presence, it informs us of the nature of the physical object and allows us to differentiate shapes, textures, materials, compliance, temperature, kinetic and other qualities.
Existing research has explored individual haptic affordances — such as changing shape, size, stiffness, and texture\cite{Hapticgoround, hapticpivot, pupop, xrings, hapticlink, pacapa} — each through separate devices designed specifically for each type of feedback.

Pneumatic inflatables are able to capture many of these affordances with their ability to dynamically inflate objects, altering their shape, size, stiffness, and texture. However, existing research on pneumatic inflatables predominantly emphasizes their role as singular passive haptic devices. Typically, these inflatables are either attached to a user’s hand \cite{pupop, Pneu-Multi-Touch:Hu} or treated as some form of prop \cite{inflatablebots, Hapticgoround, TilePoP, inflatabletextures, InflatableMod}. This limited perspective overlooks the potential of pneumatic inflatables to actively and dynamically interact with users in a more integrated and versatile manner.

To address this gap, we present Pneutouch, an untethered, wrist-worn device that supports multiple pneumatic haptic affordances, capable of delivering up to three inflatables into and out of a user's hand.
Pneutouch's combination of a dynamically programmable stiffness with its ability to deliver an inflatable into and out of a user's hand addresses this limitation by exploring the dynamic aspects of pneumatic inflatables within a user's hand. 
In this work, we explore the affordances provided by varied stiffness, user-controlled stiffness, and the active delivery of inflatables into and out of a user's hand. 

Pneutouch spotlights the potential of pneumatic inflatables in shaping diverse and immersive haptic experiences in VR. We envision that these inflatables represent a wide range of tangible virtual objects. By enabling multiple haptic affordances within a single device, Pneutouch introduces the concept of readily available shape-changing physical proxies that seamlessly enhance the tactile dimension of VR interactions. This approach emphasizes the importance of integrating multifaceted haptic feedback to enrich user experience and interaction within virtual environments.

\begin{figure}[h]
\centering   
%for sigchi
\includegraphics[width=1\columnwidth]{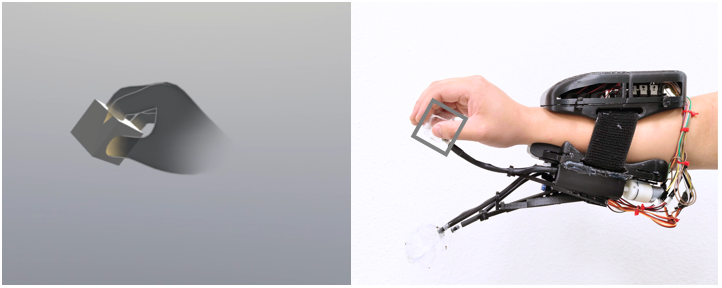}
\caption{Left: Virtual hand grabbing a cube. Right: Pneutouch delivering a cube shaped inflatable into a user's hand.}
\label{header}
\end{figure}

Pneutouch is capable of: 

\begin{itemize}
  \item Dynamic Inflation of Haptic Proxies: Dynamic inflation of inflatable proxies offers representation versatility. A single inflatable, with varying stiffness levels, can effectively portray a range of different objects — a sphere inflatable can represent the shape of a pool cue ball or a ripe tomato. A user can have programmable control over the inflatable's shape, size, and stiffness and a tangible agency in their virtual experience.

    \item Dynamically Delivered Proxies:  
    The dynamic delivery of inflatable proxies into and out of a user's hand.    
    
  \item Customizability: Easy interchangeability of inflatables allows for experimentation with diverse physical proxy designs. Tinkerers and researchers have a versatile platform to explore various haptic interactions and their impact on user perception in VR.
  
\end{itemize}

In our user studies, we evaluated the ability for pneumatic inflatables to provide haptic sensations, such as 1) changing stiffness, 2) dynamically inflating or deflating based on user input, and 3) delivering different textures to a user. We also explore contextual applications with Pneutouch, such as sorting different shaped virtual objects, target throwing, bowling and plucking shape-changing produce from a magic plant. Participants emphasized the importance of having agency in controlling object shape, feeling congruent sensations to what was seen for a more immersive experience, and being able to interact with multiple objects. Overall, Pneutouch provided users with haptic feedback that enhanced their virtual reality interactions, and improved the realism of the virtual experience.

\begin{figure*}[htbp]
    \centering
    \subfloat[Dynamic Inflation - Top row: User feeling a normal corn on the cob, fully inflated. Bottom row: User feeling a fully deflated shape changed corn]{\includegraphics[height=3.2cm]{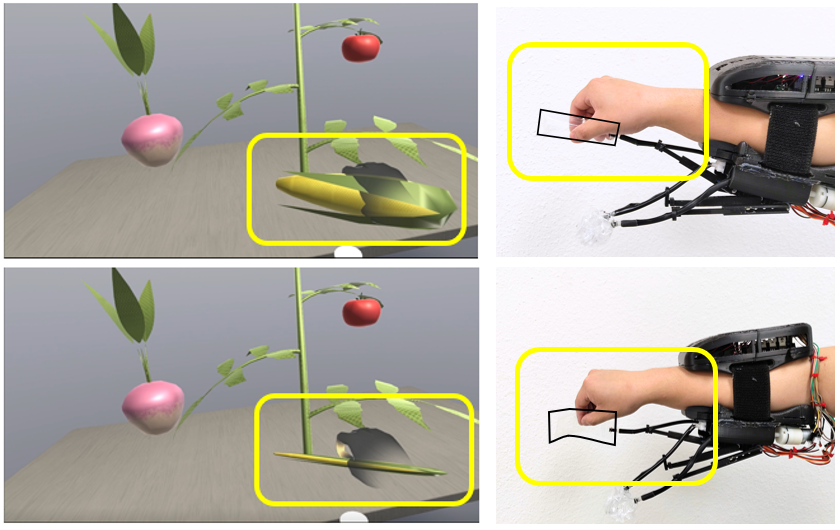}\label{fig:image1}}
    \hspace{0.3cm}%
    \subfloat[Dynamically Delivered Proxies - User throwing virtual ball. As the user releases the virtual ball, the physical inflatable ball leaves the user's hand]{\includegraphics[height=3.2cm]{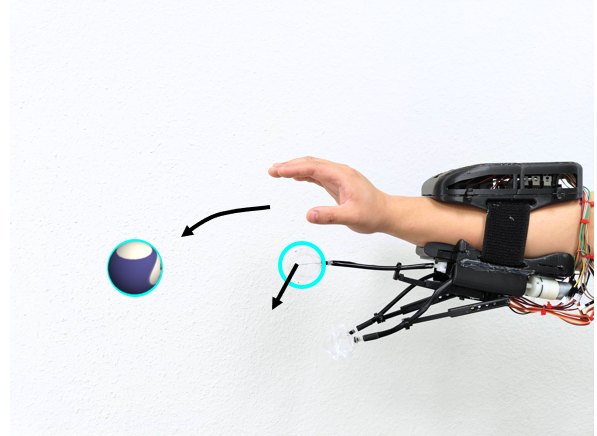}\label{fig:image2}}
    \hspace{0.3cm}%
    \subfloat[Customizable Physical Proxies - A variety of inflatables that can be attached or swapped for the Pneutouch device.]{\includegraphics[height=3.2cm]{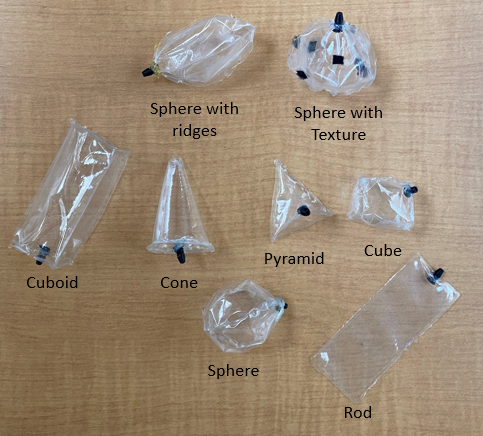}\label{fig:image3}}
    \caption{Pneutouch Capabilities}
\end{figure*}
\section{Related Works}

\subsection{Passive Haptics}
Passive haptic devices increase immersion without relying on actuation for user interaction \cite{PassiveHaptics_Insko}. The sense of touch is crucial for creating a feeling of presence, as even low-fidelity objects can enhance users' sense of presence in virtual environments \cite{VAGHELA2021249}. However, current active haptic systems that use motors to simulate tactile feedback fall short in replicating the experience of sophisticated passive haptic models \cite{VAGHELA2021249}.

Numerous works have explored the implementation of passive haptic props for improved immersion. iTurk utilizes a ceiling-tethered prop and tracking to enable real-time interaction with non-actuated haptic objects \cite{iturk}. Haptic-go-round introduces a motorized turntable that rotates the appropriate haptic device to match users' intended touch direction \cite{Hapticgoround}. Stair proxies create the illusion of walking up and down stairs \cite{stairs}. Other works, such as RoomShift \cite{Roomshift}, TouchMover \cite{TouchMover}, and Snake Charmer \cite{Snakecharmer}, utilize moving passive haptic proxies. Haptic retargeting techniques manipulate users' visual perception to reduce spatial mismatches \cite{hapticretargeting}\cite{Sparsehaptic}. Additionally, works such as HaptoBend \cite{haptobend} and Adaptic \cite{adaptic} involve shape-changing props that utilize planar panels bending along hinges.
Pneutouch's combination of wrist-worn versatility and dynamic delivery of physical proxies set it apart. 

\subsection{Wrist-worn and Handheld Haptic Devices}

In the field of haptic devices, there are wrist-worn examples that offer various types of interactions. Thermal and vibrotactile sensations on the wrist have been explored \cite{Thermohaptic:Bolton}\cite{ ThermalBracelet:Peiris}\cite{doppelheartbeathaptic:Azevedo}, along with devices delivering squeeze sensations \cite{Tasbi:Pezent}. Some wrist-worn devices utilize pneumatic actuation for the wrist and forearm. For example, Pneufetch uses pneumatically actuated nodes on a wristband to create different haptic cues \cite{Pneufetch}, and HapWRAP features inflatable tubes wrapped around the wrist to provide natural skin cues \cite{HapWRAP:Agharese}. However, these devices do not offer sensations for grasped objects.

Handheld haptic devices have also been developed to simulate grabbing and texture change. Works such as Grabity \cite{grabity} and Wolverine \cite{wolverine} employ braking mechanisms for full-hand grasping, while CLAW \cite{claw} and CapstanCrunch \cite{capstan} focus on braking mechanisms for the index and thumb. These devices, however, lack the ability to render dynamic forces of objects entering and leaving the hand. Haptic Pivot \cite{hapticpivot} addresses this by using an actuated arm to pivot a generic haptic proxy into the user's hand. Haptic Pivot is limited to a singular generic haptic proxy in a virtual reality experience and from the results of Haptic Pivot’s task 1, 5 out of the 9 virtual objects were given relatively low realism scores. Torc \cite{torc} and Haptic Revolver \cite{hapticrevolver} provide the sensation of changing textures for the fingertips.

In contrast, Pneutouch offers a broader range of haptic sensations compared to Haptic Pivot's single generic haptic proxy and other work.
Pneutouch is capable of delivering three different shaped proxies - which can be customized for texture, and shape - into the user’s hand.

Various systems have delved into handheld shape-changing devices. Xrings, for instance, employs a motor-based mechanism to achieve dynamic shape change \cite{xrings}. PoCoPo, on the other hand, utilizes a pin-based approach \cite{pocopo}. Pneutouch explores an inflatable approach for dynamic shape transformation. Notably, PuPoP, while designed for grasping, does not incorporate dynamic shape change; instead, it relies on predefined primitive shapes \cite{pupop}.

\subsection{Pneumatic Passive Haptics}

Human-computer interactions utilizing pneumatics have gained traction in various applications. Pneumatics for passive haptics have advantages of changing stiffness, shape, and size. Additionally, they are light weight and compliant. Examples include medical simulations \cite{PnematicMedicalSimulations:Talhan}. Previous works have employed pneumatic systems as haptic displays to replicate textures, such as finger top displays for robotic surgery \cite{DisplaySurgery:King} and glove-based displays \cite{WearableDisplay:Yutaka}. Volflex demonstrated the use of inflatable balloons to mimic three-dimensional objects, creating a volumetric display \cite{Volflex:Iwata}.

Pneumatic shape-changing proxies exhibit considerable promise within the realm of virtual reality. In the exploration conducted by PneUI, pneumatic inflatables were employed as shape-changing proxies, manipulating their forms through the modulation of pneumatic pressure \cite{PneuUI:Yao}. PuPoP took this concept a step further by directly attaching shape-changing inflatables to the user's hand, allowing for deflation and inflation of the objects during interactions with virtual elements \cite{pupop}.  Pneu-Multi-Tools introduced an auto-folding interface to expand the range of shapes achievable using inflatable airbags \cite{Pneu-Multi-Touch:Hu}. HaPouch employed phase change of a volatile liquid to inflate pouches on the fingertips \cite{HaPouch:Uramune}. Beyond hand-centric applications, various pneumatic interfaces have been devised for diverse purposes, including haptic feedback on the body \cite{somainflatables}, tabletop interfaces \cite{InflatableMod}, floor tiles \cite{TilePoP}, passive haptics \cite{10.1145/3472749.3474760}, and even incorporation into footwear \cite{Pushups}.

Many of these works use pneumatics as singular passive haptic devices. With our Pneutouch system, we explore the dynamic aspects of inflatables within a user’s hand. Pneutouch can also deliver up to three inflatable shaped-proxies providing multiple haptic sensations within a VR experience.

\section{Implementation}
Pneutouch's implementation comprises of a mechanical system, custom electronics, control firmware and software elements in VR. Outside of the custom PCB, the components are readily available and accessible. The total material cost for one Pneutouch device is roughly 80 dollars.

\subsection{Physical Components}

\begin{figure}[h]
\centering   
%for sigchi
\includegraphics[width=.65\columnwidth]{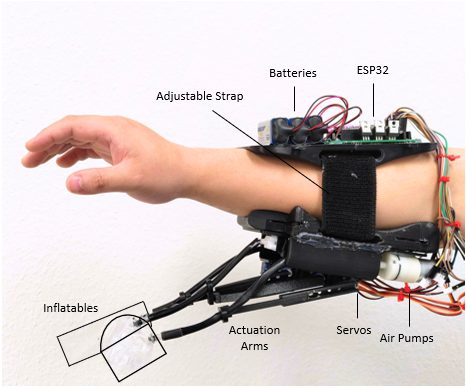}
\caption{Pneutouch functional elements, displayed without the top cover: ESP32 on PCB, actuation arms and servo motors, inflatables, air pumps, and adjustable strap (strap side behind).}

\label{fig:OnHand}
\end{figure}

Our device is completely self-contained and wireless. Figure.~\ref{fig:OnHand} shows the main components of our Pneutouch system. Our Pneutouch system weighs a total of 584g, with the three lithium ion 9v batteries weighing at 26g, and the top cover weighing 68g. The two pneumatic motors weigh 100g. The four valves weigh 48g. Three SG90 servos weigh 27g. The weight mostly comes from the PLA prints. Further refinement of the design and print settings can reduce the weight. Our device is equipped with an ESP32 2.4GHz micro-controller with built-in wi-fi connectivity. 

\begin{figure}[h]
    \centering   
    %for sigchi
    \includegraphics[width=.85\columnwidth]{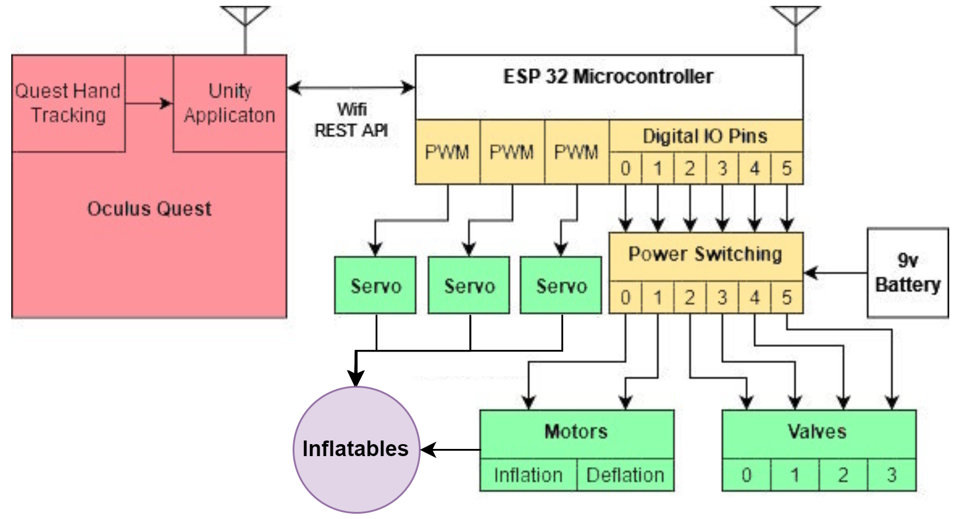}
    \caption{Pneutouch's System Schematic}
    \label{fig:SystemDiagram}
    \end{figure}

The ESP32 controls the servo motors as well as the pneumatic valves and pneumatic motors. Pneutouch uses wifi communication interface between a Meta Quest and our device. Figure.~ \ref{fig:SystemDiagram} shows our system diagram. Wifi communication allows for a wireless, low latency experience.

\subsubsection{3D Printed Components}
Pneutouch parts were designed in Autodesk Fusion360 and 3D-printed in PLA. Pneutouch is built to comfortably fit the median forearm circumference of an adult (9.3-11.7in) while remaining stationary during movement. An adjustable strap system keeps the device in place on the forearm. Arms, with adjustable length, attach to each servo motor. Clips on each arm hold the air tubing and move each inflatable into or out of the palm of the hand.  

\subsubsection{Inflatables}
The pneumatic inflatables employed in Pneutouch are made of sheet vinyl, 4mm thick, with patterns for the shape of each volume cut out by hand. The hardness of the vinyl was measured for be 19 on the Shore hardness scale. To seal the edges of the vinyl together, an impulse heat sealer\footnote{METRONIC Impulse Sealer 8 inch} was used. For different textures, we attach different material on the outside of the inflatable. The inflatables were sized accordingly as to not tangle when the arms actuated. A 3D printed tubing connection is attached to the end of each inflatable.

Air tubing runs from the air pumps to the valves and finally to each inflatable. A 1.25mm wire was securely attached to the tubing and held in place using electrical tape, allowing for easier adjustment to accommodate varying arm lengths while ensuring the tubing maintains a rigid shape.

For our user studies, we decided to implement simple shapes, such as a cube, a sphere and a rod-shaped inflatable. These common abstract shapes are effective analogs to many objects that one would interact with in virtual spaces. These shapes took roughly 0.2 seconds to go from fully deflated to fully inflated, and roughly 1 second to go from fully inflated to fully deflated. Each inflatable, taking roughly 0.2 seconds to inflate, has a volume of approximately 0.008334 liters.

\subsubsection{Electronics}
We use two 4.5 V pneumatic DC motors\footnote{Adafruit Product ID: 4700} from Adafruit, one pump to inflate and one pump to deflate. The flow rate of each pneumatic motor is 2.5 liters per minute and has a 'stall' max pressure of -55 kpa. Each pump has a draw of 500mA. We use four 6V air valves\footnote{Adafruit Product ID: 4663} with 2-pin JST PH Connectors from Adafruit. These air valves serve as pneumatic relays. We use silicone tubing to transfer air from the pneumatic motors, through the valves and into the inflatables.

\begin{figure}[h]
\centering   
%for sigchi
\includegraphics[width=.8\columnwidth]{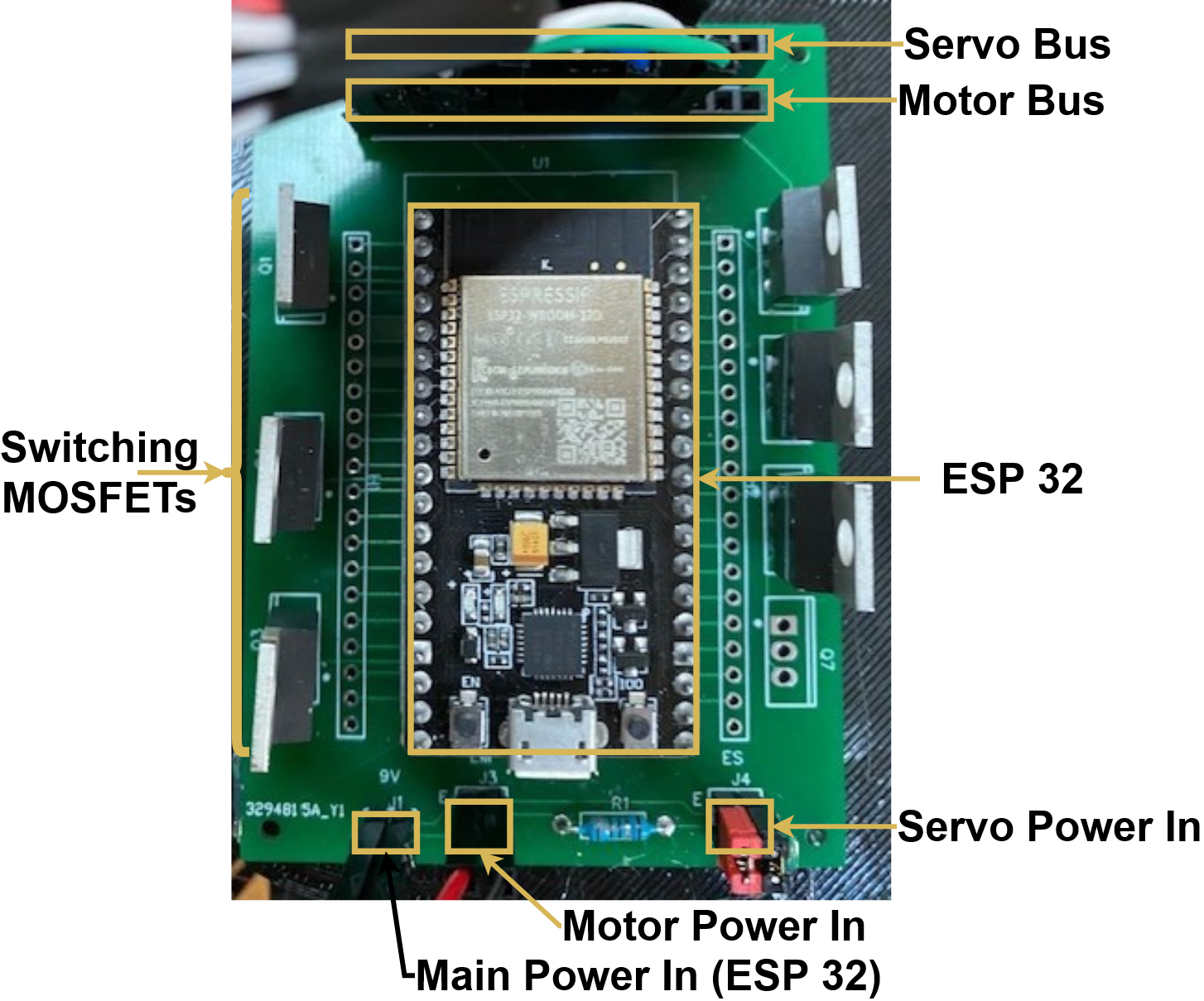}
\caption{PCB and Electronic Components}
\label{fig:PCB}
\end{figure}

The active electronic components of the Pneutouch system include six IRLB8721 MOSFETs and the ESP32 microcontroller. All seven components are connected together on a printed circuit board (PCB) that was designed using an electronic design automation (EDA) software. The PCB also links together all of the passive electronic components in the circuit which include three micro servo motors, two pneumatic air pumps, and four air valves. 
Figure~\ref{fig:PCB} shows the electronic components in our custom PCB.

The PCB is designed so that the ESP32 sits in the center with three MOSFETs on either longitudinal side. Each MOSFET is connected at its gate node to one pin on the ESP32. There are two rows of female pin header sockets at the back end of the PCB which serve as the connection points for the air valves, air pumps, and servo motors. One row has pins to the drain of the six MOSFETs. This row is used specifically for the air valves and air pumps so that they can be dynamically switched on and off. Each air valve and air pump is connected to one MOSFET and when the ESP32 sends a signal to the MOSFET, it allows current to flow which turns on the corresponding air valve or air pump. The MOSFETs are rated for a max of 30V drain-to-source voltage, but the air valves and air pumps never use more than 5V which keeps the MOSFETs deep in the triode region of operation. The second row of pin header sockets is used for the three servo motors. These pins connect the servo motors directly to the main power source as well as three digital pins on the ESP32. Using the ESP32, the servo motors can be controlled to rotate to any position needed. Without any load, we measured the SG90 servo motors to rotate 60 degrees in 0.9 +/- 0.1 seconds. The arm would travel roughly 60 degrees to deliver the inflatable from ready to grasping angle into the palm of the user. The PCB also includes ports for the main power source.

\subsection{Software Control}
For tracking the user's hand, we use the Meta Quest's hand tracking API. Our Pneutouch system does not require any external controllers or trackers to calculate the position of the hands. Our Pneutouch system does not hinder or occlude the hand tracking.

To control the pneumatic valves, pneumatic motors and servo motors, we wrote our own custom APIs. The  API is agnostic of hardware and integrable into numerous software solutions. The software can be split into two sections, a REST server running on a microcontroller, and a Unity interface. We chose a REST server over other options due to its ease of integration in various game platform and relatively low latency. 

\subsubsection{Microcontroller API}
The REST Server runs locally over wifi on an ESP 32 microcontroller and exposes two GET endpoints. 

\verb|/setBatch?pin=<string of 2 digit pin IDs>|

\verb|&state=<digital state of corresponding pin 0/1>|, and
\verb|/setServo?pin=<pin ID>|
\verb|&state=<angle 0-180>|. \verb |setBatch| allows for dynamic control of all digital IO pins on the microcontroller that could correspond to various hardware configurations. For example, if pin 19 should be on, and 18 off on the microcontroller, the command would follow: \verb|/setBatch?pin=1918&state=10|. In the request, the first state “1” corresponds to the first two digit pin ID “19” and the second state “0” corresponds to the second pin ID “18”.   Likewise, \verb|/setServo| allows for the control of any servo connected to a PWM pin. 

%For example, if the servo attached to PWM (Pulse Width Modulation) pin 4 should be set to 90 degrees, the following command can be sent: \verb |/setServo?pin=4&state=45|. Together, both endpoint give full access to the microcontroller hardware and allows specific actions to be abstracted to the game engine side, creating a hardware, configuration agnostic solution.

We took 10,000 measurements of the setServo and setBatch commands. We found that the setServo latency took on average 0.023 seconds with a standard deviation of 0.088 seconds. We found that for the setBatch latency took on average 0.019 seconds with a standard deviation of 0.033 seconds.

\begin{figure}[h]
\centering   
%for sigchi
\includegraphics[width=1\columnwidth]{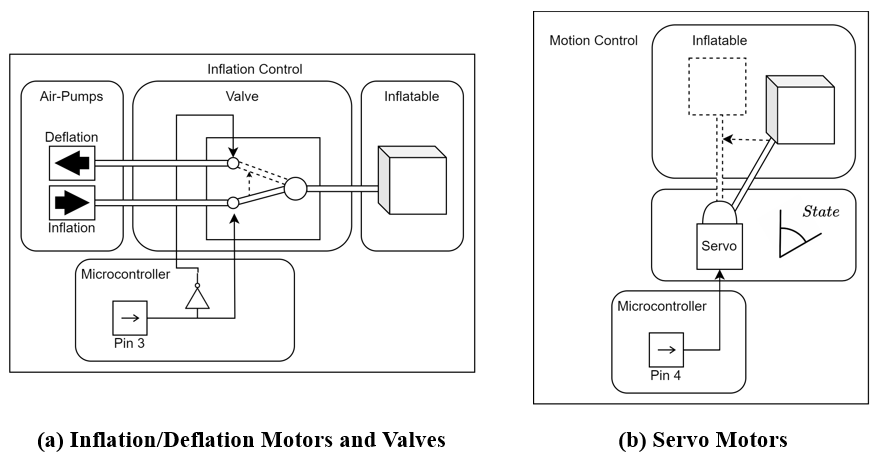}
\caption{Hardware implementation diagram of the inflation and servo mechanisms}
\label{fig:simpleSystem}
\end{figure} 

\subsubsection{Unity Game Engine API}
We split the game engine API into two sections; one controls the inflation and deflation of objects, and the other controls the servo location of those objects. We utilize the Meta Quest hand tracking within the Unity Game Engine.

Inflation and deflation utilizes Unity’s collider system to register when the player’s hand approaches the virtual object. Each virtual object has a \textit{PneuObject} component that holds the interaction name and registers the collision. Once a collision occurs, \textit{PneuObject} invokes the \textit{PneuStatic} object to set the microcontroller state. \textit{PneuStatic} is a static class that stores the combination of pins on and off, which is required for inflation. \textit{PneuStatic} passes the state combination to a \textit{Network Controller} object that interfaces with microcontroller REST endpoints setting the desired state. This is done utilizing the \verb|/setBatch| endpoint. For instance, an inflation command like \verb|/setBatch?pin=010203&state=100| triggers inflation by activating the specified pins. This state persists until the player withdraws their hand, at which point PneuStatic inverts the states for deflation.
 
%For example, say a hardware configuration is as follows: (see figure \ref{fig:simpleSystem} ) inflation motor pin 1, deflation motor pin 2, and valve pin 3 (0 state is inflation). For inflation the following command would be sent \verb|/setBatch?pin=010203&state=100|, starting the inflation motor, setting the deflation motor off, and setting the valve to inflation. The inflation state runs until the user moves their hand away from the virtual object. Once the user’s removes their hand from the collider the process happens again, but instead PneuStatic inverts the appropriate states (as specified by the de-signer) to invoke deflation. 

Additionally, two specific modes enhance this functionality. Variable inflation allows designers to specify inflation levels from 0 to 1 for an object. When interacted with, PneuObject inflates the object for a designated time to achieve the desired inflation level. User input inflation is based off hand squeeze. When the user squeezes, the system responsively inflates or deflates the object. The squeeze is characterized by the distance between the tip and knuckle index finger joint when the hand squeezes. These modes are easily toggled in the Unity Game Engine.

%Augmenting this standard behavior are two PneuObject specific modes. Variable Inflation, allows the designer to specify a range of inflations from 0 no inflation to 1 max inflation for a specific object. Then when interacted with the PneuObject invokes inflation for a specific amount of time to obtain the desired inflation level. The other mode is, user input inflation, in this mode the PneuObject tracks how much the user is squeezing their hand and responsively deflates or inflates the object accordingly. Hand squeeze is characterized through the tracking of specific finger joints that are articulated when the user squeezes an object with their hand. Specifically, the tip and the knuckle of the index finger are tracked, and the distance between the two joints is calculated. In both cases the enabling of this mode of operation is easily accomplished through toggling of a Boolean in the Unity Game Engine.

%Using the example above, the design would specify pin 1, 2, and 3 as pins to invert, because the motors must flip operation (inflation stop, deflation start) and the valve must switch which motor drives the inflatable. Therefore, to deflate the command \verb|/setBatch?pin=010203&state=011| would be sent. Having the effect of turning off the inflation motor, turning on the deflation motor, and switching the valve to deflation. The deflation command runs for a designer specified amount of time, which can vary depending on the size of the inflatable. Which is not limited to a set number of inflation or predefined actions.  

To adjust the servo position, a REST API call is made to the ESP microcontroller, specifying either the ready angle or the grasping angle. The servo control in Pneutouch operates based on distance between hand and the virtual objects. When the user's hand is within a delta distance from the virtual object the servo is triggered to actuate into the grasping radius. This delta distance is manually set by the developer. We set our delta distance was set at 1cm after we tested many different distances. We found that 1cm was the best balance of responsiveness and accommodating the natural movement of the user's hand in the virtual space.

Each virtual object can be assigned to an individual servo motor. If an object is within the ready radius of the user's hand, the servo initializes to a ready angle and adds the object to the list of "ready objects." The list is checked to ensure that only one object is presented to the user's hand at a time. If there are multiple objects in the grab range then it will calculate which is closer based on distance and then put the closer one in the grabbed state.

\section{Core Device Attributes User Study}
In this user study, our primary objective was to delve into the core attributes of Pneutouch that set it apart from existing research work. We conducted a user study to evaluate Pneutouch’s ability to simulate: (1) variable stiffness (2) dynamic inflation in response to user input and (3) variable textures for pneumatic inflatables. We recruited 14 participants aged 21 to 35: 6 identified as male, 6 as female, 1 as non-binary, and 1 preferred not to disclose their gender. All 14 participants had previous experience with VR, 11 of which had prior experience with haptic feedback. All participants were predominantly right-handed. Only right handed interactions were done. All studies were reviewed and approved by the institutional review board.

In these studies, we hypothesized that our Pneutouch system with its dynamic inflation and deliverance of haptic proxies would provide more congruent sensations in variable stiffness, dynamic inflation in response to user input and variable texture compared to similar research work that only focused on one of these affordances. Our participants played three unique minigames utilizing three input interfaces. The first input system was using Pneutouch on ``Pivot Only" where only the servo arm actuated and there was a spherical foam ball attached that acted as the generic haptic proxy. The second input interface was using Pneutouch on ``Inflation Only" where only the pneumatic motors actuated and a spherical inflatable was stuck onto an wristband the user wore on the palm. The third interface was using our normal Pneutouch system which was ``Inflation and Pivot".
The combination of ``Inflation and Pivot" offers opportunities to explore new interactions.
Meanwhile, the ``Pivot Only" and ``Inflation Only" are different modalities that Pneutouch support inspired by HapticPivot \cite{hapticpivot} (188g) and PuPop \cite{pupop} (less than 10g). We made our best attempt to accurately replicate device conditions from the literature for a system to system comparison. The order of the inputs were randomly selected for each minigame.

\begin{figure}[h]
\centering   
%for sigchi
\includegraphics[width=.95\columnwidth]{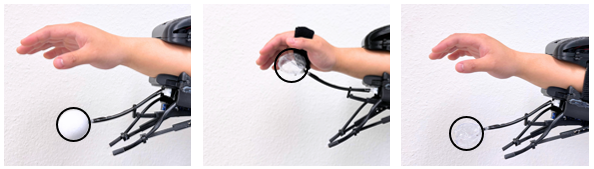}
\caption{Pivot Only, Inflation Only, and Inflation and Pivot interfaces}
\label{modes}
\end{figure}

After each input interface for each minigame, participants were asked to provide their feedback on their individual experiences through a series of questions where users rated the enjoyment and object realism using a 7-point likert scale. They were also asked open-ended questions: 1) How did what they see correspond to what they felt in the [minigame specific interaction], 2) Describe their experience with each interface, and 3) How did [minigame specific interaction] impact their experience \footnote{List of study questions can be found in supplemental material}. After each minigame, users had the option to take a 5 minute break. 

For analysis of the data, we conducted the Friedman test to determine whether there were statistically significant differences among the groups for both enjoyment and object realism scores. If there were sigificant differences, we then used the Nemenyi post-hoc test to find significant differences between all group pairs. We also share notable user responses.

\subsection{Variable Stiffness: 4 Balls of Different Stiffness}

Variable inflation allows for the creation of a spectrum of tactile sensations, ranging from soft and gentle touches to firm, robust interactions. In the first minigame, we measured Pneutouch's ability to render virtual objects of varying stiffness. Within the virtual scene, four different spheres would appear on a table before the user. Each of these spheres had different inflation levels of max (16 PSI), medium (8 PSI), low (1.6 PSI) and min (0 PSI). The virtual spheres were identical in size to ensure an unbiased exploration of haptic feedback. We asked participants to reach towards and grab the virtual spheres. The users were asked to do this for roughly a minute for the three input interfaces (Pivot Only, Inflation Only and Inflation and Pivot).

\begin{figure}[h]
\centering   
%for sigchi
\includegraphics[width=.95\columnwidth]{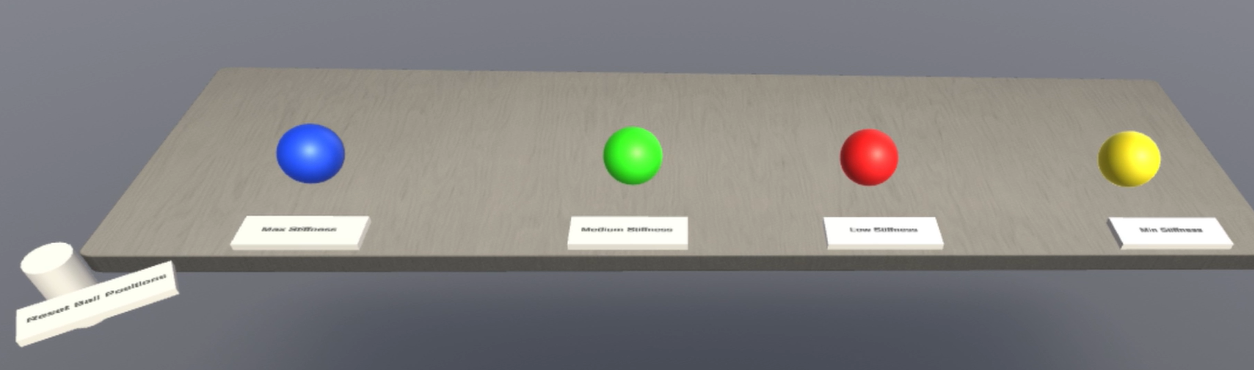}
\caption{Spheres at varying stiffness levels: max, medium, low and min.}
\label{variablestudy}
\end{figure}

\textit{Results}. In the variable stiffness minigame, no significant differences were found among the three interfaces with the Friedman test for both enjoyment and realism. Users rated Pivot Only with mean enjoyment and object realism scores of 5.85 (SD = 0.95) and 4.93 (SD = 1.63), respectively. Inflation Only received ratings of 5.36 (SD = 0.92) for enjoyment and 4.29 (SD = 1.43) for object realism. Users rated Inflation and Pivot with mean enjoyment and object realism scores of 5.64 (SD = 0.63) and 4.71 (SD = 1.20), respectively. 

Two users noted that they expected the virtual ball to "squish" according to its stiffness. The visual consistency of the ball may have caused confusion, as it didn't change shape. For instance, two different users rated “Pivot Only” a 7 in object realism because it matched their visual expectations. We believe this mismatch may have influenced the overall ratings.

\begin{figure}[h]
\centering   
%for sigchi
\includegraphics[width=0.95\columnwidth]{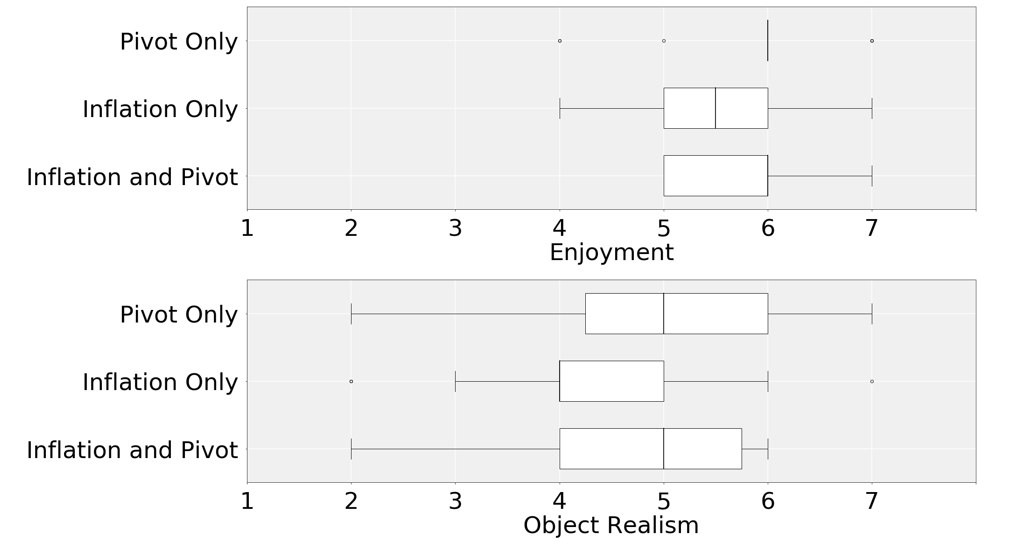}
\caption{Variable inflation study results, user ratings for enjoyment and object realism. Pivot Only was the Haptic Pivot stand-in. Inflation Only was the PuPop stand-in. Inflation and Pivot was using our Pneutouch system normally.}
\label{fig:userstudy4}
\end{figure}

The qualitative responses of the users showed that ``Inflation and Pivot" outperformed ``Inflation Only" and ``Pivot Only" in providing a more immersive and varied experience of stiffness. The combination of both pivot and inflation mechanisms in ``Pivot and Inflation" seemed to offer a more nuanced representation of stiffness, as users found it easier to distinguish levels and appreciated the correlation between physical sensations and visual representations. 12 of the users were able to feel different stiffness with this interface. User 13 writes, ``Response time was phenomenal, I think this and combination of a more reactive stiffness feature, makes the experience more realistic, compared to just the inflatable or just the pivot." 2 users did report that visually the virtual spheres were slightly larger than what they felt.

The ``Pivot Only" interface received mixed reviews from users, with challenges reported in distinguishing distinct levels of stiffness using this method alone. 10 of the users reported feeling the same stiffness while 4 thought they might have felt some differences but were not sure. User 10 shares, ``I feel the hardness of the physical ball's surface is different, but I'm not sure because the difference is not very obvious to me." 

User feedback for the ``Inflation Only" interface also presented a varied response. 10 users were able to feel different stiffness with this interface. Users reported challenges for grabbing the physical inflatable. Participants generally noted the impact of inflation on the physical ball, with comments from User 13, such as ``Yes, with the inflation I definitely felt noticeable levels of stiffness. I can tell that the highest stiffness was more inflated in my hand, and the lowest stiffness was almost completely deflated." Three users highlighted challenges in grabbing the physical inflatable in this mode. Given that the inflatable was slightly smaller than what was seen virtually, User 9 describes, ``Because this was inflation only, I had to reach my fingers down to find/feel the inflatable in real life. This meant the physical world didn't perfectly align with what I was experiencing in VR, which somewhat broke the illusion."

\subsection{Dynamic Inflation in Response to User Input}

By allowing virtual shapes to dynamically change based on user actions, a deeper level of agency is introduced into the virtual experience. In the third minigame we explore this concept and we measured Pneutouch's ability to dynamically deliver air to an inflatable in response to the user's input. Users had control of the physical inflatable proxy. For example, when the user squeezed slightly, the inflatable would begin to fully deflate. When the user stopped squeezing, the inflatable would revert back to full inflation (16 PSI). The users were asked to do this for roughly a minute for the three input interfaces (Pivot Only, Inflation Only and Inflation and Pivot).

\begin{figure}[h]
\centering   
%for sigchi
\includegraphics[width=0.95\columnwidth]{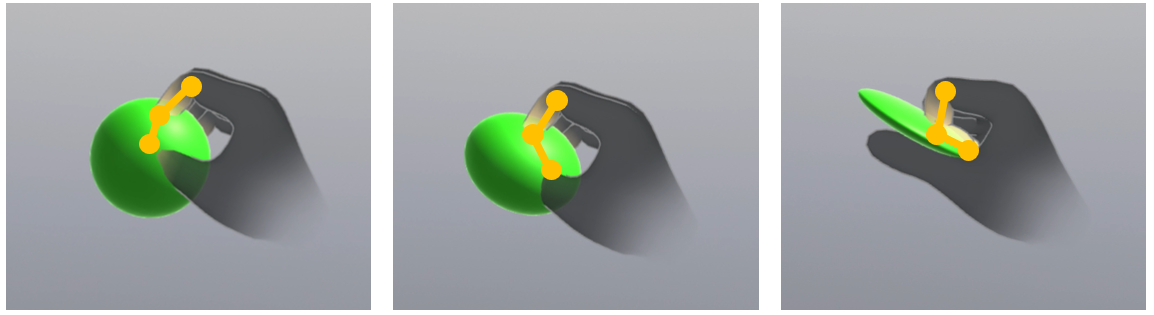}
\caption{Stiffness of object based on user hand pose. The more the user curls the index finger, the more deflated the haptic proxy becomes.}
\label{userinputstudy}
\end{figure}

\textit{Results}. For the dynamic inflation based on user input, the Pneutouch mode of Inflation and Pivot was rated the highest among the three interfaces for enjoyment (M = 5.92, SD = 0.73) and object realism (M=6, SD=0.96) compared to the Inflation Only (enjoyment: M =5.42, SD=1.15; object realism: M = 5.42, SD=1.01) and Pivot Only (enjoyment: M=4.43, SD=1.09; object realism:2.86, 1.41).

\begin{figure}[h]
\centering   
%for sigchi
\includegraphics[width=0.95\columnwidth]{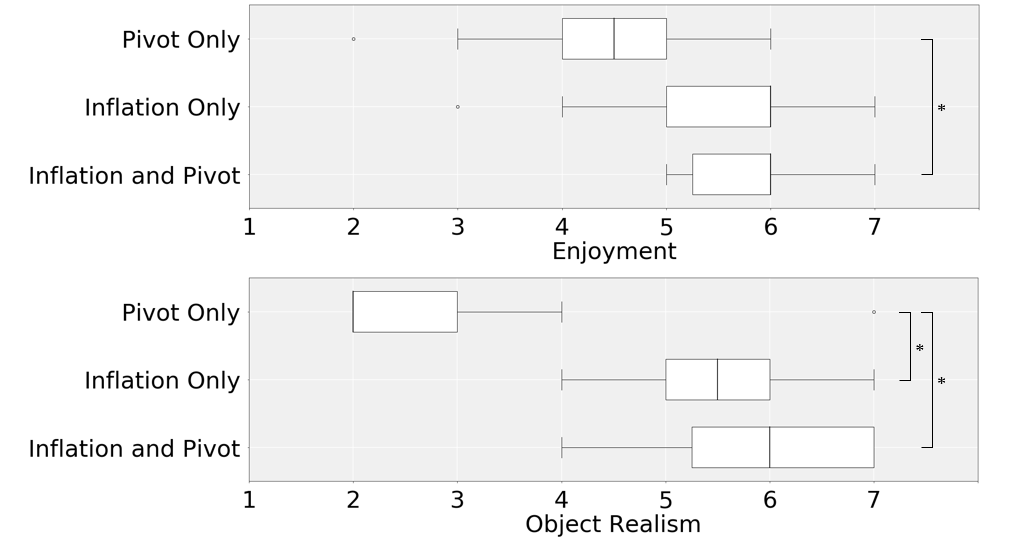}
\caption{Dynamic inflation in response to user input study results, user ratings for enjoyment and object realism. Pivot Only was the Haptic Pivot stand-in. Inflation Only was the PuPop stand-in. Inflation and Pivot was using our Pneutouch system normally}
\label{fig:userstudy6}
\end{figure}

Significant differences were observed between Inflation and Pivot compared to Pivot Only for both enjoyment (P \textless 0.01) and object realism (P \textless 0.01).
Additionally, statistically significant differences were found between Inflation Only and Pivot Only for object realism (P \textless 0.01).

``Pivot and Inflation" emerged as the most effective and immersive interface. Users reported positive experiences with the combination of both pivot and inflation mechanisms. This approach allowed users to dynamically change the virtual objects by both squeezing the ball and adjusting their finger positions, providing a more natural and responsive interaction. User 11 shares, ``My hand was empty when my hands were in the air, I could freely use them for other things. When I grabbed the ball, it appeared in my hand and according to the pressure I applied, it deflated. This experience was much better than the previous versions of the game."

``Inflation Only" was rated similarly in enjoyment and object realism, as users also enjoyed being able to squeeze and cause the object to inflate and deflate. 2 users noted that they had trouble picking objects up. User 8 mentions, ``I felt more in control of the system. It was definitely a little hard picking up the ball and getting the ball to inflate and deflate, but it felt pretty accurate when it worked "

``Pivot only" faced the most significant challenges, with users struggling to effectively change the virtual objects and experiencing limitations due to the physical styrofoam ball. The styrofoam ball often obstructed the desired shape-changing actions and users were unhappy not being able to deflate the ball. User 5 writes, ``As I tried to squish the object, the pivot ball got in the way and prevented me from squishing as easily from the last example."

\subsection{Variable Texture: a Stone Block, a Kiwi and a Starfruit}
Feeling textures in VR enhances the immersive experience by providing users with feedback that corresponds to the virtual objects. In the second minigame, we measured Pneutouch's ability to render virtual objects of varying textures. Within the virtual scene, users were able to pick up a stone block, a kiwi and a starfruit. Rough tape was attached on the outside of the inflatable for the stone block, brown felt for the kiwi, masking tape on the edge for the starfruit. The users were asked to do this for roughly a minute for the three input interfaces (Pivot Only, Inflation Only and Inflation and Pivot). With Inflation and Pivot, we had three different haptic inflatables with different textures. With the inflation only, the kiwi textured inflatable was attached to the wristband around the user's palm as PuPop only had one inflatable per experience. In the pivot only mode, the generic spherical foam ball was used.

\begin{figure}[h]
\centering   
%for sigchi
\includegraphics[width=.95\columnwidth]{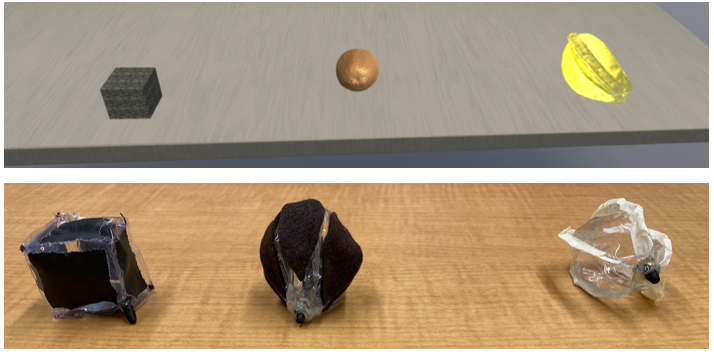}
\caption{Top row: Virtual objects of rough block, kiwi and starfruit. Bottom row: Textured physical inflatables with corresponding textures to match the virtual objects.}
\label{texturedstudy}
\end{figure}

\textit{Results}. Participants using the Inflation and Pivot (Normal Pneutouch) reported higher levels of enjoyment (M=6.14, SD=0.77) and object realism (M=6, SD=0.87) compared to Inflation Only (enjoyment: M =4.71, SD=1.06; object realism: M=3.92, SD=1.43) and Pivot Only (enjoyment: M=5.07, SD=1.13; object realism: M=3.5, SD=1.29).

\begin{figure}[h]
\centering   
%for sigchi
\includegraphics[width=0.95\columnwidth]{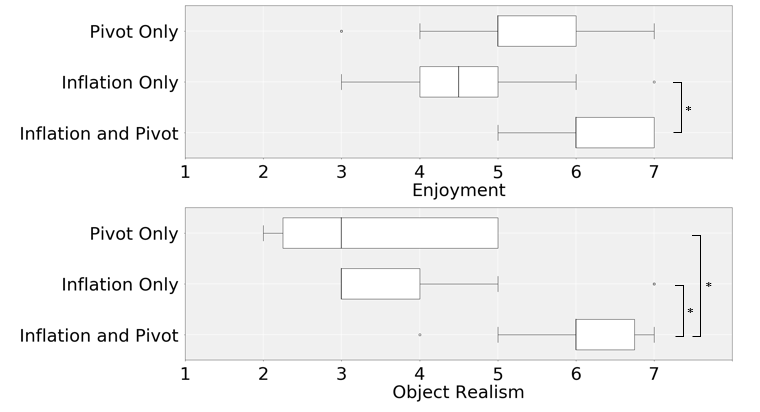}
\caption{Texture study results, user ratings for enjoyment and object realism. Pivot Only was the Haptic Pivot stand-in. Inflation Only was the PuPop stand-in. Inflation and Pivot was using our Pneutouch system normally}
\label{fig:userstudy5}
\end{figure}

Participants using the Pneutouch mode of Inflation and Pivot reported \textit{significantly} higher enjoyment scores compared to Inflation Only (p \textless 0.01).
Participants using the Pneutouch mode of Inflation and Pivot also rated \textit{significantly} higher object realism scores compared to Inflation Only (p \textless 0.03), and Pivot Only (p \textless 0.01).

In direct comparison to ``Inflation Only" and ``Pivot Only," the ``Inflation and Pivot" interface outperformed due to its ability to provide a more comprehensive and realistic haptic experience. Users consistently reported feeling distinct textures and shapes, enhancing immersion and engagement. The synchronized feedback from both pivot and inflation mechanisms created a more cohesive and convincing simulation, aligning physical and virtual sensations effectively. In the ``Pivot and Inflation" condition, user 4 highlights, ``The texture as well as the stiffness felt right especially between the kiwi and starfruit, the loosey feel of the starfruit was nice and the still feel of the kiwi was also nice". User 3 adds, ``Definitely felt the difference. The iron block felt like something hard is being grabbed".

The main issue users had with the ``Inflation Only" and ``Pivot Only" interfaces was that they were only able to feel a single texture that did not match what they saw visually. The the respective papers, Haptic Pivot used a generic proxy without textures and PuPop only had a single inflatable in the user’s hand that could not be switched. In the ``Inflation Only", user 12 highlights, ``it was almost misleading to touch the same texture." In the ``Pivot Only", user 14 shares, ``It wasn't pleasant. Without textures, it made me more aware that I was in VR and this wasn't real." 
\section{Contextual Device Applications User Study}

In this user study, we aim to assess contextual device applications using Pneutouch. We developed four unique minigames where we evaluated Pneutouch’s ability to simulate: (1) grasping virtual objects of varying shapes, (2) picking up and throwing overhand (3) picking up and throwing underhand/sideways, (4) shape changing virtual objects for pneumatic inflatables. We recruited twelve participants (7 male, 5 female) between ages 20-33. 3 users had never used VR before. 7 of the users who had used VR had not used VR with any sort of haptic feedback. All participants were right-handed. Only right handed interactions were done. All studies were reviewed and approved by the institutional review board.

We hypothesized that our Pneutouch system would provide more congruent haptic sensations in the contextual applications compared to hand tracking, and quest controllers that vibrated on object pickup.
The hands and controllers were baseline conditions.
The order of the inputs were randomly selected for each minigame.

After each input interface for each minigame, participants were asked to provide their feedback on their individual experiences through a series of questions where users rated the enjoyment and object realism using a 7-point likert scale. They were also asked open-ended questions: 1) Describe their experience for each interface, 2) Did what they see correspond to what was felt, why or why not, and 3) How did haptic feedback if any impact their experience \footnote{List of study questions can be found in supplemental material}. After each minigame, users had the option to take a 5 minute break. 

For analysis of the data, we conducted the Friedman test to determine whether there were statistically significant differences among the groups for both enjoyment and object realism scores. If there were sigificant differences, we then used the Nemenyi post-hoc test to find significant differences between all group pairs. Insights from user responses are described in the discussion section.

\subsection{Sorting Objects: Grabbing Objects of Varying Shapes}
\begin{figure}[h]
\centering   
%for sigchi
\includegraphics[width=0.95\columnwidth]{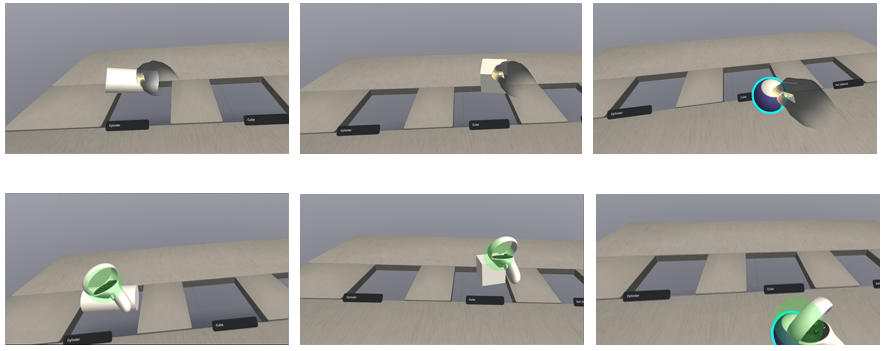}
\caption{Top row: User grasps cylinder, cube, and sphere with right hand. Bottom row: Same objects grabbed with right controller. In Hand and Pneutouch, user sees top row; in Controller, bottom row view.}
\label{userstudy1}
\end{figure}

In the first minigame, we measured Pneutouch's ability to render the sensation of grasping different virtual objects (cylinder, cube, and sphere). Within the virtual scene, a cylinder, cube or sphere would appear on a table before the user. We asked participants to reach towards and grab the virtual object and place it into its respective hole (Figure~\ref{userstudy1}). The users were asked to do this for roughly a minute for the three input interfaces (Pneutouch, hands and controller).

\textit{Results.} 
Participants reported higher enjoyment (M = 5.5, SD = 1.44) and perceived object realism (M = 5.16, SD = 1.33) when using Pneutouch to grab virtual objects of varying shapes compared to using their hand (enjoyment: M = 4.58, SD = 1.67; object realism: M = 2.08, SD = 1.83) or a controller (enjoyment: M = 5.17, SD = 0.83; object realism: M = 3.25, SD = 1.42).

Participants using Pneutouch reported \textit{significantly} higher object realism than compared to using hands (P \textless 0.01).

\subsection{Target Throwing: Rendering Feedback of Overhand Throws}

\begin{figure}[h]
\centering   
%for sigchi
\includegraphics[width=0.95\columnwidth]{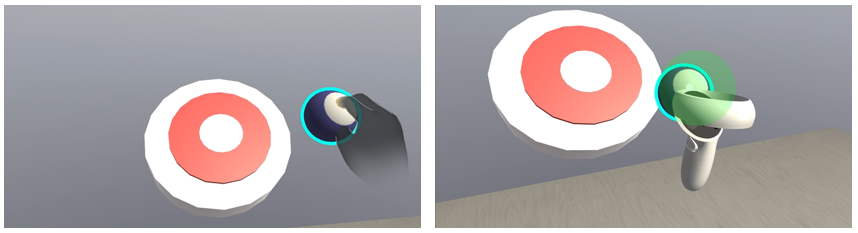}
\caption{Left: User throws cue ball at target with right hand. Right: Throws cue ball at target with right controller. In Hands and Pneutouch, scene is on left; in Controller, on right.}
\label{userstudy2}
\end{figure}

In the second minigame, we measured Pneutouch's ability to dynamically deliver proxies when picking up and releasing a virtual object when the object was thrown overhand. Within the virtual scene, a sphere would appear on a table before the user. We asked participants to reach towards and grab the virtual object and throw it at the target (Figure~\ref{userstudy2}). The users were asked to do this for roughly a minute for the three input interfaces (Pneutouch, hands and controller).

\textit{Results.} 
Participants reported higher enjoyment (M = 5.58, SD = 1.78) and perceived object realism (M = 5.58, SD = 1.24) when using Pneutouch to throw a virtual object overhand at a target compared to using their hand (enjoyment: M = 4, SD = 1.95; object realism: M = 2.58, SD = 1.72) or a controller (enjoyment: M = 5.16, SD = 1.26; object realism: M = 3.33, SD = 1.26). 

Participants using Pneutouch reported \textit{significantly} higher object realism than compared to using hands(P \textless 0.01) or compared to using the controller(P=0.038).

\subsection{Cue Ball Bowling: Rendering Feedback of Sideways/Underhand Throws}

\begin{figure}[h]
\centering   
%for sigchi
\includegraphics[width=0.7\columnwidth]{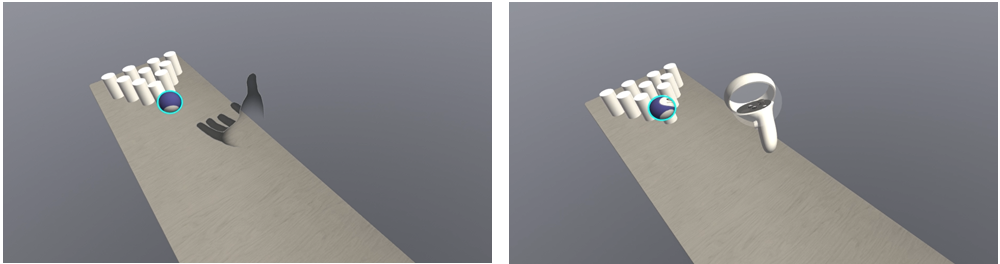}
\caption{Left: User throws cue ball at pins with right hand. Right: User throws cue ball at pins with right controller. In Hands and Pneutouch, user sees scene as on left; in Controller, as on right.}
\label{userstudy3}
\end{figure}

In the third minigame, we measured Pneutouch's ability to dynamically deliver proxies when picking up and releasing a virtual object when the object was thrown side ways or underhand. Within the virtual scene, a cue ball would appear on a table before the user. We asked participants to reach and grab the virtual object and throw it either sideways or underhand at the virtual pins (Figure~\ref{userstudy3}). The users were asked to do this for roughly a minute for the three input interfaces (Pneutouch, hands and controller). The user could press the white button (not pictured) near the front of the table to reset the pins and the sphere.

\textit{Results.} 
Participants using Pneutouch reported comparable levels of enjoyment (M = 5.5, SD = 1.24) and higher perceived object realism (M = 5.91, SD = 1.78) when throwing a virtual object sideways/underhand to knock over pins, compared to using their hand (enjoyment: M = 4.9, SD = 1.5; object realism: M = 3.16, SD = 1.89) or a controller (enjoyment: M = 5.66, SD = 1.07; object realism: M = 4, SD = 1.71). 

Participants using Pneutouch reported \textit{significantly} higher object realism than compared to using hands(P \textless 0.01).

\subsection{Magic Garden: Shape Changing Fruits and Vegetables}

\begin{figure}[h]
\centering   
%for sigchi
\includegraphics[width=0.95\columnwidth]{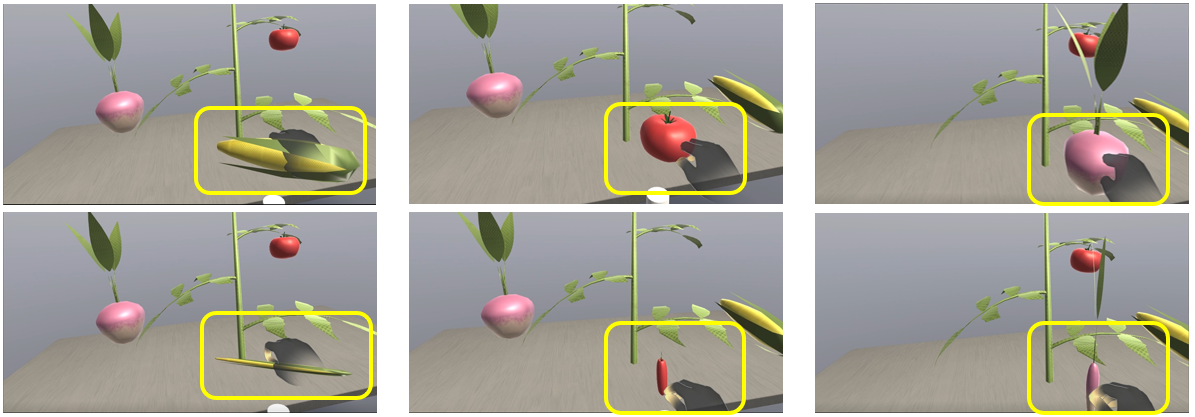}
\caption{Top row: User holds virtual produce in original shape. Bottom row: Same virtual item, shape-changed. In Hands and Pneutouch, user sees hand; in Controller (not pictured), a controller is shown.}
\label{userstudy4}
\end{figure}

In the fourth minigame, we measured Pneutouch's ability to render shape change of virtual objects. Within the virtual scene, there was a virtual plant with different produce growing from it. We asked participants to reach towards and pluck a fruit or vegetable from the plant and then tap it to the white button. Plucking the produce caused the servo arm to actuate 2 degrees down to simulate the plucking force. The produce would change shape when the button was pressed. The produce would either shrink from normal to skinny, or grow from skinny to normal as shown in Figure~\ref{userstudy4}. The users were asked to do this for roughly a minute for the three input interfaces (Pneutouch, hands and controller).

\textit{Results.}
Pneutouch outperformed in interacting with multiple shape-changing virtual objects in the garden scene. Participants using Pneutouch reported significantly higher enjoyment and object realism compared to using their hands (Enjoyment: P=0.038, Object Realism: P=0.038) or using the controller (Enjoyment: P=0.002, Object Realism: P=0.016). 

Pneutouch also had notably higher average enjoyment (M=6.25, SD=1.14) and object realism scores (M=6.09, SD=1.24) compared to the hand condition (Enjoyment: M=4, SD=1.81, Object Realism: M=2.25, SD=2.094) and the controller condition (Enjoyment: M=4.33, SD=1.67, Object Realism: M=2.74, SD=1.82).

\section{Discussion}

\subsection{Dynamic Inflation - Programmable Stiffness and Shape Change}

In our exploration of dynamic inflation for programmable stiffness and shapes through user studies involving objects of variable stiffness and dynamically changing shapes, we discovered its pivotal role in offering a versatile and immersive haptic experience. This significance became particularly apparent in scenarios where programmable stiffness and dynamic shape changes were crucial. The ability to dynamically adjust inflation levels based on user input empowered users not only to perceive changes in pressure and stiffness but also to actively control and shape virtual objects by manipulating their physical counterparts. The programmable nature of dynamic inflation provide new opportunities for interactions in VR.

\subsection{Combination of Affordances Better Than One}
In comparing our Pneutouch device as ``Inflation and Pivot" with PuPop \cite{pupop} as ``Inflation Only" and Haptic Pivot \cite{hapticpivot} as ``Pivot Only", Pneutouch's combination of dynamic inflation  and deliverance of multiple proxies excelled in conveying varying stiffness, varying texture, and dynamic inflation based on user input.

The limitation of offering a single haptic proxy in both ``Inflation Only" and ``Pivot Only" interfaces had a noticeable impact on the overall user experience, hindering the versatility and richness of haptic feedback in virtual reality (VR) interactions. In the ``Inflation Only" interface, the band along the palm presented some issues where the some users had trouble grabbing the physical inflatable. Additionally, the band may not allow a user to feel the full texture of what's been picked up. Within the texture minigame, with ``Inflation Only" we had the kiwi inflatable attached to the palm. This may have been a limitation as there might have been different results with a different textured inflatable. In the ``Pivot Only" interface, the use of a generic haptic proxy, represented by the physical styrofoam ball, presented several challenges that impacted the overall user experience negatively. The generic nature of the haptic proxy, which did not adapt or convey specific variations in virtual objects, led to a lack of precision, realism, and versatility in haptic feedback.

While the combination of affordances is better than one, there is a trade off of having more affordances and interactions to weight. The Pneutouch system at 584g weighs more than HapticPivot (188g) and PuPop (less than 10g). Future iterations of Pneutouch can take on the challenge of having a combination of affordances with a lighter design and smaller, more powerful components.

\subsection{Contextual Device Applications}
Our contextual application studies revealed insights into the impact of haptic feedback on user experience. Across all studies, users consistently emphasized the importance of tactile sensations for immersion and realism. Pneutouch received positive feedback for enhancing enjoyment and realism. Users appreciated feeling the shape of objects, which increased their sense of control and engagement. In contrast, conditions lacking haptic feedback, such as using controllers or hands alone, resulted in diminished experiences, with users expressing dissatisfaction due to the absence of tactile sensations and realism in interactions with virtual objects.

In activities like target throwing and bowling, where the physical sensation of picking up and releasing objects is crucial, Pneutouch outperformed other conditions by providing accurate feedback that enhanced gameplay enjoyment. Similarly, in the shape change scenario, Pneutouch excelled in providing precise haptic feedback for interactions like plucking fruits and vegetables and experiencing shape changes, leading to a more immersive and enjoyable experience compared to conditions lacking such feedback. 
User feedback highlight that traditional hands and controller input systems are insufficient in providing tactile experiences for users.

\section{Limitations}

\subsection{User Study Sample Size}
In our user studies we had sample sizes of 14 and 12 users. Larger sample size studies could enhance the statistical power to detect significant differences and improve the ability to generalize the findings.

\subsection{Meta Quest Hand Tracking and Control}
Our implementation of the Pneutouch device relies on the functionality of Meta Quest hand tracking. Any issues with tracking can disrupt the device's performance, causing unexpected behavior like incorrect arm movements or failure to inflate/deflate the inflatables. Users have reported issues with control, such as virtual objects not behaving as expected or virtual hands getting stuck until hands become visible again. Users also took some time learning to grab virtual objects with hand tracking.

\subsection{Physical Limitations}

Rapid arm movements occasionally caused the device to slip, such as in the bowling study. As a result, the arms with the inflatables attached became misaligned with the palm, and the inflatables failed to reach the user's hand effectively. In the testimonials, 4 users found the weight of the Pneutouch device on their forearms to be a little heavy. Improvements to the physical design would help lighten the system when worn and prevent any unwanted sliding. We want to minimize the amount of 3D printed material to house the electronics. This can be done through lightweighting, strategically placing holes in the design to maintain structural integrity while minimizing material usage. 

Users also commented on the noise generated by the pneumatic motors. While the majority of users were able to overlook these sounds, three users mentioned in the testimonials that they found the noise to be distracting and slightly disorienting. We can explore the possibility of incorporating quieter and faster motors to mitigate the impact of the motor sounds or having users wear noise canceling headphones.

\subsection{Physical Inflatable Proxies}

Feedback and observations from users highlighted several limitations of the Pneutouch inflatable proxies, including orientation, weight, tangling, and inflation/deflation speeds.
One user experienced discomfort due to a mismatch in orientation caused by the rod. Using all shapes with symmetric volumes, or adding additional degrees of freedom to the mechanism, could prevent such issues in future iterations.
Although Pneutouch effectively delivers the inflatables as physical proxies, they primarily convey shape rather than weight. Future work can explore having the servo arm exert downward force to simulate a sense of weight when the user grabs the inflatable.
To avoid the arms from getting caught on each other or tangling, the size of the inflatables was kept small in our implementation. Using inflatables that incorporate origami principles or utilizing two actuating arms could allow for the use of larger inflatables.
Testimonials from user studies showed that most users found the shape morphing from inflation and deflation to be quick and responsive. 3 users thought the inflation and deflation speeds could be faster.
Faster inflation/deflation could be achieved by using more powerful motors or employing visuo-haptic illusions to make the visual transformation process coherent with the inflatable's state in the hand.
\section{Future Work}
\textbf{Exploring Visuo-haptic illusions}
In a related work, the authors of Pupop \cite{pupop} utilized visuo-haptic illusion to represent four different sizes using a single prop. Future work could explore the application of visuo-haptic illusions to represent a broader range of sizes for multiple inflatable props. Additional work can also explore visual-haptic stiffness, building on \cite{pseudostiffness}'s research on pseudo-stiffness in non-compressible objects. Moreover, investigating the impact of object orientation or mismatches when placed in a user's hand could offer valuable insights into enhancing VR experiences. 

\textbf{Advanced Inflatables} We plan to explore various manufacturing techniques that enable prop extension, shape stacking, shape change, and the creation of complex shapes. We plan to investigate the implementation of methods proposed by Pneuseries\cite{Pneuseries} or Siloseam\cite{siloseam}. Additionally, we intend to explore the utilization of capacitive touch sensing \cite{capacitivetouch} to transform the inflatables into input devices.

\textbf{Alternate Uses} Our vision for the Pneutouch system goes beyond the conventional haptic interactions. For example, Pneutouch might simulate playing virtual bongos. When the user's hand contacts the bongo at different spots, a proxy could simulate the impact, varying in stiffness. Another application involves integrating different worn inflatables\cite{compressables} with the Pneutouch device. This approach opens doors to immersive haptic experiences. Imagine a virtual trip to the doctor's office: an inflatable armband inflates and deflates, replicating a checkup procedure. The creative potential of Pneutouch offers opportunities to enhance the virtual reality landscape.

\section{Conclusion}
We explored the affordances and interactions of pneumatic inflatables with our Pneutouch system that delivered readily available shape changing physical proxies into and out of the user's palm. We demonstrated the potential of Pneutouch as a platform for enhancing haptic feedback in virtual environments. From user ratings and testimonials, the combination of "Inflation and Pivot" afforded by Pneutouch outperformed a single affordance such as "Inflation Only" and "Pivot Only". Pneutouch also provided a more enjoyable and realistic experience compared to hands and controllers. Users highlighted Pneutouch's ability to emulate varying stiffness, responsiveness to gestures, and texture. Pneutouch opens new avenues of haptic feedback exploration within virtual reality experiences.

\bibliographystyle{plain}
\bibliography{pneutouch}
\end{document}